\documentclass[a4paper,11pt]{article}
\usepackage{pos}
\newcommand{\rev}[1]{ {\color{black}{#1}}}
\DeclareMathOperator{\SU}{\mathrm{SU}}
\DeclareMathOperator{\Tr}{\mathrm{Tr}}

\newcommand{\redchisq}{\chi^2_{\tiny\mbox{red}}}

\newcommand{\SW}{S_{\mbox{\tiny{W}}}}

\newcommand{\eq}{\begin{equation}}
\newcommand{\en}{\end{equation}}
\newcommand{\bra}{\left \langle}
\newcommand{\ket}{\right \rangle}

\title{On the behaviour of the interquark potential in the vicinity of the deconfinement transition}

\author[a]{F. Caristo}
\author*[a,b]{M. Caselle}
\author[c]{N. Magnoli}
\author[a]{A. Nada}
\author[a,b]{M. Panero}
\author[a]{A. Smecca}

\affiliation[a]{Department of Physics, University of Turin \& INFN, Turin,\\
 Via Pietro Giuria 1, I-10125 Turin, Italy}

\affiliation[b]{Arnold-Regge Center, University of Turin,\\
Via Pietro Giuria 1, I-10125 Turin, Italy}

\affiliation[c]{Department of Physics, University of Genoa \& INFN, Genoa,\\
Via Dodecaneso 33, I-16146 Genoa, Italy}

\emailAdd{caselle@to.infn.it}

\abstract{In the vicinity of the deconfinement transition the behaviour of the interquark potential can be precisely predicted using the Effective String Theory (EST). If the transition is continuous we can combine EST results with a conformal perturbation analysis and reach the degree of precision  needed to detect the corrections beyond the Nambu-Got{\={o}} approximation in the EST.  We discuss in detail this issue in the case of the deconfinement transition of the $\SU(2)$ gauge theory in $(2+1)$ dimensions (which belongs to the same universality class of the 2d Ising model) by means of an extensive set of high precision simulations.
 We show that the Polyakov loops correlator of the $\SU(2)$ model is precisely described by the spin-spin correlator of the 2d Ising model not only at the critical point, but also  
 down to temperatures of the order of $0.8 T_c$. 
 Thanks to the exact integrability of the Ising model we can extend the comparison in the whole range of Polyakov loop separations, even beyond the conformal perturbation regime. 
We use these results to quantify the first EST correction beyond  Nambu-Got{\={o}} and show that it is compatible with the bounds imposed by a bootstrap analysis of EST. This correction encodes important physical information and may shed light on the nature of the flux tube and of its EST description.}

\FullConference{%
 The 38th International Symposium on Lattice Field Theory, LATTICE2021
  26th-30th July, 2021
  Zoom/Gather@Massachusetts Institute of Technology
}

\begin{document}
\maketitle

\section{Introduction}

\label{sec:introduction}

In the last few years there has been a lot of progress in the description of the confining regime of Yang-Mills theories using the  ``Effective String Theory'' (EST) approach in which the confining flux tube joining a quark-antiquark pair is modeled as a thin vibrating string~\cite{Nambu:1974zg,Goto:1971ce,Luscher:1980ac,Luscher:1980fr,Polchinski:1991ax} .  
In particular it has been realized that the EST enjoys the so called ``low energy universality'' property~\cite{Luscher:2004ib,Aharony:2013ipa}: due to the peculiar features of the string action and to the symmetry constraints imposed by the Poincar\'e invariance in the target space, the first few terms of its large distance  expansion are fixed and coincide with those obtained terms in the same expansion for the Nambu-Got{\={o}} string. This implies that the EST is much more predictive than other effective theories and in fact its predictions have been be successfully compared in the past few years with many results on the interquark potential from Monte Carlo simulations in Lattice Gauge Theories (LGTs).
However it is clear that the Nambu-Got{\={o}} action should be considered only as a first order approximation of the actual EST describing the non-perturbative behaviour of the Yang-Mills theories. Going beyond this approximation is one of the most interesting open problems in this context.
It turns out that the optimal setting to address this issue is the finite temperature regime of the theory, in the neighbourhood of the deconfinement transition, but still in the confining phase since in this regime the boundary corrections become subleading and can be neglected \cite{Caselle:2021eir}.

In this contribution we report the results, that we recently discussed in ref.~\cite{Caristo:2021tbk}, of a set of simulations of the $\SU(2)$ gauge theory in $(2+1)$ dimensions. 
We studied the Polyakov loop correlators in the range of temperatures $0.8 T_c\leq T \leq T_c$ (where $T_c$ denotes the deconfinement temperature) from which we extracted the interquark potential to be compared with the effective string predictions.
Since the deconfinement transition for this model is of the second order, simple renormalization group arguments (the so called ``Svetitsky-Yaffe conjecture'' \cite{Svetitsky:1982gs}) show that in the neighbourhood of the deconfinement transition the model is in the same universality class of the bidimensional Ising model.  Thus the Polyakov loop correlators from which we extracted the interquark potential can be mapped using this correspondence to the spin-spin correlators of the $2d$ Ising Model which, thanks to the exact integrability of the model, are exactly known.
We shall see that this mapping is impressively confirmed by the simulation and that the knowledge of the exact behaviour of the correlator will help us to extract with very high precision the temperature dependence of the interquark potential and, from that, the corrections with respect to the Nambu-Got{\={o}} action which is the main goal of our study.

\section{Lattice setup and definitions}
\label{sec:setup}
 We simulated the $\SU(2)$ model on a (2+1) dimensional cubic lattice of spacing $a$ and sizes $N_t$ in the compactified ``time'' direction (which plays, as usual, the role of inverse temperature $T=1/N_t$) and $N_s$ in the two other (``spatial'') directions using the standard Wilson action
\begin{equation}
\SW = -\frac{\beta}{2} \sum_{x} \sum_{0 \le \mu < \nu \le 2} \Tr U_{\mu\nu} (x)
\end{equation}
where $U_{\mu\nu}(x)$ is the plaquette operator and $\beta$ is related, as usual, to the coupling constant of the model: $\beta=4/(ag^2)$.
To simplify notations we set the lattice spacing $a=1$ and neglect it in the following.
One of the advantages of studying the $\SU(2)$ model in (2+1) dimensions is that
we can leverage previous studies to fix the parameters of the model.  In
particular we can use the scale setting expression obtained in
\cite{Teper:1998te}   
\begin{equation}\label{scale_setting}
\sqrt{\sigma_0(\beta)}=\frac{1.324(12)}{\beta}+\frac{1.20(11)}{\beta^2}+\mathcal{O}(\beta^{-3})\ ,
\end{equation}
 where $\sigma_0$ denotes the zero temperature string tension.
Moreover some of the simulations that we shall discuss were performed at 
$\beta=9.0$ for which an independent determination of $\sigma_0$ can be found in \cite{Bonati:2021vbc} and \cite{Caselle:2004er}.
{A set of high precision simulations at $\beta=16.0$ can be found in \cite{Athenodorou:2016kpd}.}
For the critical temperature we shall use the estimates of  \cite{Edwards:2009qw}.

We shall be interested in the following in the  Polyakov loops correlator 
\begin{equation}
\label{def_G}
G(R) = \left\langle \sum_{\vec{x}} \ P\left(\vec{x}\right) P\left(\vec{x}+R \hat{k}\right) \right\rangle ,\hskip 1cm
P\left(\vec{x}\right) = \frac{1}{2} \Tr \prod_{t=0}^{N_t} U_0 \left(t,\vec{x}\right).
\end{equation}
where $\hat{k}$ denotes one of the two ``spatial'' directions, the sum is over all spatial coordinates $\vec{x}$ and
the Polyakov loop $P\left(\vec{x}\right)$ at a spatial coordinate $\vec{x}$ is defined as the trace of the closed Wilson line in the compactified ``time'' direction. In a  finite temperature setting the interquark potential $V(R,N_t)$ can be extracted from the free energy associated to the 
Polyakov loop correlator:
\eq
G(R) ~\equiv ~ {\rm e}^{-\frac{1}{T}V(R,N_t)}~ = ~{\rm e}^{- N_t V(R,N_t)}~~,
\label{polya}
\en
\noindent
In the confining phase we expect, for large values of $R$, a linearly rising potential $V(R,N_t)=\sigma(T) R$ where $\sigma(T)$ denotes the {\sl finite temperature} string tension. As we shall see below $\sigma(T)$ is a decreasing function of $T$ and vanishes exactly at the deconfinement point. From $V(R,N_t)$ and $\sigma(T)$ we can also define the zero temperature potential $V(R)$ as the $T\to 0$ limit of $V(R,T)$ and the zero temperature string tension $\sigma_0$ as the $T\to 0$ limit of $\sigma(T)$.

\section{The ``Svetitsky-Yaffe'' mapping}
\label{sysec}
It is well know that a powerful way to study the behaviour of a Yang-Mills theories in the neighbourhood of the deconfinement transition consists in building an effective action for the Polyakov loops, by integrating out the spacelike links of the model. 
Starting from a ($d+1$) dimensional LGT we end up in this way with an effective action for the Polyakov loops which will be a $d$ dimensional spin model, having a global symmetry with respect to the center of the original gauge group.
If the deconfinement transition is continuous, in the vicinity of the critical point the fine details of the effective action can be neglected, and it can be shown~\cite{Svetitsky:1982gs} that it belongs to the same universality class of the simplest spin model, with only nearest neighbour interactions, sharing the same symmetry breaking pattern. 
This means, in our case, that the deconfinement transition of the $\SU(2)$ LGT in $(2+1)$ dimensions, which is continuous, belongs to the same universality class of the symmetry breaking phase transition of the two-dimensional (2d) Ising model. In particular the Polyakov loop correlator in the confining phase is mapped to the spin-spin correlator of the 2d Ising model in the symmetric, high-temperature phase of the model,  for which  an exact expression is known~\cite{Wu:1975mw} and can be expanded in the short distance ($R\ll\xi$) and in the large distance ($R\gg\xi$) limits (where $\xi$ denotes the correlation length)  as follows:

\begin{itemize}
\item 
{\bf $R\ll\xi$}

\begin{equation} \label{shortR}
    \langle{\sigma(0)\sigma(R)}\rangle = \frac{k_s}{R^\frac{1}{4}}\bigg[1+\frac{t}{2} \ln\bigg(\frac{e^{\gamma_E}t}{8}\bigg)+\frac{1}{16}t^2+\frac{1}{32}t^3\ln\bigg(\frac{e^{\gamma_E}t}{8}\bigg)+O(t^4\ln^2t)\bigg],
\end{equation}

where $t=\frac{R}{\xi}$, ${\gamma_E}$ denotes the Euler $\gamma$ constant while $k_s$ is a non-universal constant, which can be evaluated exactly in the case of the 2d Ising model on a square lattice \cite{Wu:1975mw}.

\item 
{\bf $R\gg\xi$}

\begin{equation} \label{largeR}
    \langle{\sigma(0)\sigma(R)}\rangle = {k_l} K_0(t),
\end{equation}
where,  again, $k_l$ is a non-universal constant (related in a non-trivial way to $k_s$) and $K_0$ is the modified Bessel function of order zero, whose large distance expansion is
$K_0(t)\sim \sqrt{\frac{\pi}{2t}}e^{-t}$.

\end{itemize}

\section{Effective String Theory predictions}
\label{sec:est}
In a finite temperature setting it is possible to show, with very mild assumptions \cite{Luscher:2004ib}, that the EST description of the confining flux tube implies the following form for the Polyakov Loop correlator:
\begin{equation} \label{EST}
    \bra{P(0)P^\dagger(R)}\ket = \sum_n |v_n(N_t)|^22R\bigg(\frac{E_n}{2\pi R}\bigg)^\frac{D-1}{2}K_{\frac{D-3}{2}}(E_nR)
\end{equation}
where $D$ denotes the number of spacetime dimensions (in our case $D=3$), $E_n$ are the energy levels of the string and $v_n(N_t)$ their amplitudes, which in general depend on the inverse temperature $N_t$.

In the particular case of the Nambu-Got{\={o}} EST it is possible to give an explicit expression for the energy levels $E_n$ which are:
\begin{equation}
  {E}_n=\sigma_0 N_t
  \sqrt{1+\frac{8\pi}{\sigma_0 N_t^2}\left[-\frac{1}{24}\left(D-2\right)+n\right]}.
\label{energylevels}
\end{equation}

For the $D=3$ case that we are interested in, the lowest state  is 
\eq
E_0=\sigma_0 N_t
  \sqrt{1-\frac{\pi}{3\sigma_0 N_t^2}} \equiv \sigma(T) N_t ,
\label{E0}
\en  
where we introduced the Nambu-Got{\=o} prediction for the finite temperature string tension $\sigma(T)$.
As we have seen in the previous section $E_0$ is the inverse of the correlation length,  thus we see from the above expression that the Nambu-Got{\={o}} EST predicts a critical temperature 
$T_{c,NG}=\sqrt{\frac{3\sigma_0}{\pi(D-2)}}$
and a critical index $\nu=1/2$. We see from the Svetitsky-Yaffe mapping (which for our model implies the 2d Ising value $\nu=1$) that the EST prediction cannot be correct.
In addition, also the EST prediction for the critical temperature is wrong (albeit close to the correct one).
These observations suggest the need to go beyond the Nambu-Got{\={o}} approximation.
In the large-$R$ limit  the Polyakov loop correlator is dominated by the lowest state $E_0$ and the corrections that we are studying appear as deviations in the expansion of $E_0$ in powers of $1/N_t$, with respect to the Nambu-Got{\={o}} result.
It can be shown that,  due to the so called ``low-energy universality'',  the first non-universal term beyond the Nambu-Got{\={o}} action can appear, in the high-temperature regime which we are studying here, only at order $1/N_t^7$.  Remarkably enough, also this correction is not completely free.  Using the notations of refs.~\cite{EliasMiro:2019kyf,EliasMiro:2021nul} this additional term can be written as
 \eq
 -\frac{32\pi^6}{225}\frac{\gamma_3}{\sigma^3N_t^7},
 \label{gamma3}
 \en
 where $\gamma_3$ is a new parameter which, together with $\sigma_0$, defines the EST. By using 
 a bootstrap analysis it is possible to show that $\gamma_3$ is constrained to be $\gamma_3>-\frac{1}{768}$.

\section{Simulation results}
We performed two sets of simulations. In the first one, which was mainly devoted to a test of the Svetitsky-Yaffe mapping, we fixed a few values of $N_t$, namely $N_t=6,7,8,9$ and varied the temperature by changing $\beta$ in the range $0.8< T/T_c< 1$. In the second one, which was devoted to a high-precision study of the correction to the Nambu-Got{\={o}} action, we chose the opposite strategy and fixed three values of $\beta$ and varied the temperature by changing $N_t$. Further details about the simulations can be found in ref.~\cite{Caristo:2021tbk}.

\begin{figure}[!htb]
  \centering
  \includegraphics[width=0.95\textwidth, clip]{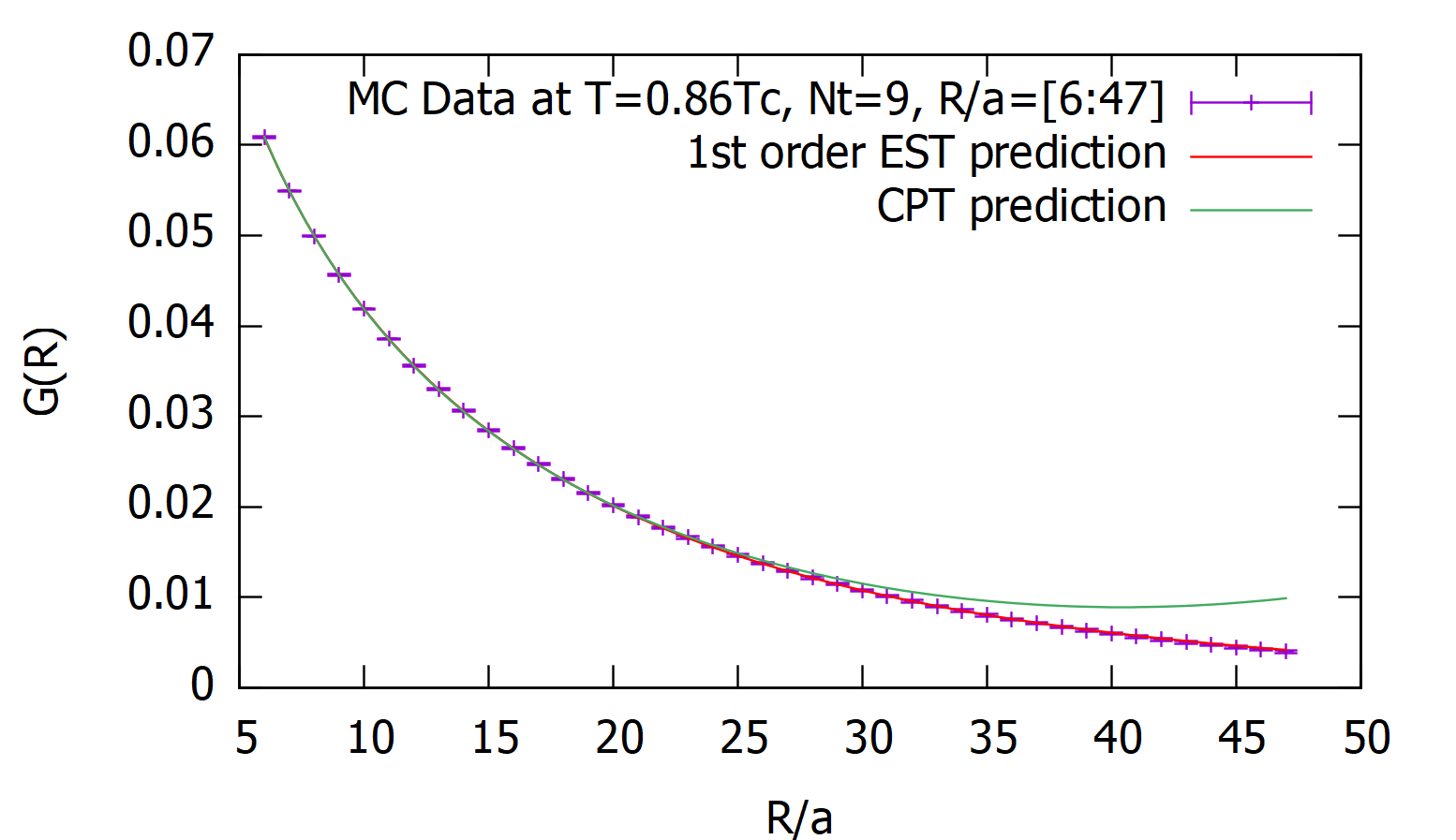}
  \caption{Fit of the data at $N_t=9$, $\beta=12.15266$, $T/T_c=0.86$ combining both the short distance approximation~(\ref{shortR}) and the large-distance one~(\ref{largeR}) for the spin-spin Ising correlator, taking into account the periodic boundary conditions.}\label{fig_3}
\end{figure}

The results of our first set of simulations show that the Svetitsky-Yaffe mapping is almost exact in the range of temperature that we tested and that there is an impressive agreement between the Polyakov loop correlators and the Ising predictions of eqs.~(\ref{shortR}) and~(\ref{largeR}).
This agreement is well visible in fig.~\ref{fig_3}, where we plotted the fit of $G(R)$ for one particular choice of $\beta$ and $N_t$ with the Ising predictions of eqs.~(\ref{shortR}) and~(\ref{largeR}) (modified to account for the periodic boundary conditions). We found a similar agreement for all the values of $\beta$ and $N_t$ that we tested.

In the second set of simulations we studied the behaviour of the ground state $E_0$ as a function of the temperature, as the deconfinement transition is approached from below, at a fixed value of $\beta$ and varying the temperature by changing the value of $N_t$. We studied the three values of $\beta$ and the range of values of $N_t$ shown in tab.~\ref{table:3}.
For each of these simulations we extracted the value of $E_0$ using the large-distance fit of eq.~(\ref{largeR}).
{We report in this proceeding a preliminary analysis of these data. Further details and a larger set of simulations can be found in
~\cite{Caristo:2021tbk}.}

The Nambu-Got{\={o}} expectation for $E_0$ is given by eq.~(\ref{E0}), which can be rewritten in terms of the critical temperature as follows:
\begin{equation} 
   E_0 = N_t\sigma(N_t)  = N_t\sigma_0\sqrt{1-\frac{\pi}{3N_t^2\sigma_0}} = N_t\sigma_0\sqrt{1-\frac{T^2}{T_{c,NG}^2}}.
   \label{NG}
\end{equation}
It is clear that this equation cannot agree with the data since it predicts a mean-field-like critical index $\nu=1/2$ for the correlation length $\xi=1/E_0$,  while the Svetitsky-Yaffe mapping predicts $\nu=1$. The latter type of scaling could be realized in several ways.  The simplest possibility is to assume a linear behaviour for $E_0$,
\begin{equation} \label{linear}
    E_0 \sim 1-\frac{N_{t,c}}{N_t}.
\end{equation}
On the other hand, the low energy universality would suggest instead an expression (using the parametrization of 
refs.~\cite{EliasMiro:2019kyf,EliasMiro:2021nul}) of the type:
\begin{equation} \label{universality}
    E_0(N_t)=\mbox{Taylor}_4(E_0)+\frac{k_{4}}{(\sigma_0)^3N_t^7}
\end{equation}
with
\begin{equation} 
 \mbox{Taylor}_4(E_0) \, \equiv \, \sigma_0 N_t - \frac{\pi}{6N_t}-\frac{\pi^2}{72(\sigma_0)N_t^3}-\frac{\pi^3}{432(\sigma_0)^2N_t^5}
 -\frac{5\pi^4}{10368(\sigma_0)^3N_t^7},
\label{Taylor}
\end{equation}
where we are truncating the Taylor expansion of eq.~(\ref{NG}) to the $1/N_t^7$ order and we assume that all higher orders are negligible, i.e. that terms arising from higher order terms of eq.~(\ref{NG}) and subleading corrections beyond Nambu-Got{\={o}}, which are not known yet, are randomly distributed and, on average, tend to cancel against each other.

Following the above observations we first tried to fit the data with a pure Nambu-Got{\={o}} ansatz $(\nu=1/2)$ and the pure linear ansatz $(\nu=1)$ of eq.~(\ref{linear}),  but neither choice provided a good fit to the data.  
 It is easy to see that the Nambu-Got{\={o}} ansatz fits well the data for larger values of $N_t$ (lower temperatures), but as the deconfinement transition is approached, it misses the approach to the critical point.  Similarly, the naive linear fit of eq.~(\ref{linear}) agrees with the data near the critical point, as expected from the Svetitsky-Yaffe correspondence,  but this agreement holds only for the first few values of $N_t$.  For larger values the naive linear  function is not compatible with our simulation results.
On the contrary the low energy universality ansatz of eq.~(\ref{universality}) turned out to fit the data remarkably well in the whole range of values of $N_t$ and for the three values of $\beta$: the results are reported in  tab.~\ref{table:3}.

\begin{table}[h]
\centering
    \begin{tabular}{ |c|c|c|c|c|c|c| }
    \hline
    $\beta$ & $N_{t,\mbox{\tiny{min}}}$ & $N_{t,\mbox{\tiny{max}}}$ & $k_4$ & $\sigma_0$ & $\redchisq$ & literature 
    \\ \hline \hline
    $9$ & $6$ & $12$ & \rev{$0.040(8)$} & \rev{$0.02603(19)$} & \rev{$1.60$} & $0.02583(3)$ \\
    \hline
    $12.15266$ & $8$ & $14$ & \rev{$0.054(5)$} & \rev{$0.01366(5)$} & \rev{$0.89$} & $0.01371(29)$\\
    \hline
    $13.42445$ & $9$ & $15$ & \rev{$0.053(8)$} & \rev{$0.01104(5)$} & \rev{$1.33$} & $0.01108(23)$ \\
    \hline
    \end{tabular}
    \caption{Results of the fit performed with eq.~(\ref{universality}). In the last column we report the values of $\sigma_0$ quoted in ref.~\cite{Bonati:2021vbc} for $\beta=9$ and in ref.~\cite{Teper:1998te} for the other two values of $\beta$.}
    \label{table:3}
\end{table}

It is interesting to note that the quality of the fits improves as we approach the continuum limit.  Moreover a highly non-trivial consistency check of the procedure is  that the best fit values for $\sigma_0$ agree in all three cases with the known ones. 
Another non-trivial check of our analysis is that the three values of $k_4$ that we found should be compatible with each other, by virtue of the $1/\sigma_0^3$ normalization of eq.~(\ref{universality}). Tab.~\ref{table:3} confirms that this is indeed the case and that the three values almost agree within the errors.

{The values of $k_4$ quoted in tab.\ref{table:3} should be considered only as preliminary estimates, since, due to the large values of the correlation length, larger values $N_s$ (the lattice size in the spacelike directions) should be tested to be sure that finite size effects are under control. While we are unable for the moment to give a precise estimate of the error on $k_4$ we quote, as  a preliminary, qualitative, estimate for $k_4$, the
value $k_4 \sim 0.05$ from which we can extract, using eq.~(\ref{gamma3}),
\eq
\gamma_3=-\frac{225}{32\pi^6} k_4 \sim - 0.0004
\en
which is well inside the bound $\gamma_3\geq-\frac{1}{768}\sim -0.0013$ obtained in ref.~\cite{EliasMiro:2019kyf,EliasMiro:2021nul}.

It is reassuring the fact that a similar analysis performed on the data of \cite{Athenodorou:2016kpd} leads to values (depending on the 
order at which the Taylor expansion is truncated) in the range $k_4\in[0.06,0.08]$, which are very close to our preliminary estimate.}

The fact that $\gamma_3$ is negative is rather non-trivial: in particular, as shown in ref.~\cite{EliasMiro:2019kyf}, it does not allow to prove the so called ``Axionic String Ansatz'' \cite{Dubovsky:2013gi,Dubovsky:2014fma}, for the EST describing this gauge theory.

It is interesting to compare this value, with the one obtained in ref.~\cite{Dubovsky:2014fma}  for the $\SU(6)$ theory in (2+1) dimensions using the data from ref.~\cite{Athenodorou:2011rx}, which is of similar magnitude but opposite in sign.  This shows explicitly that at this level of resolution the EST is not universal anymore but encodes, as expected, the specific properties of the underlying Yang-Mills theory.

Finally, it is important to stress that our results are not limited to the $\SU(2)$/Ising mapping and to the fact that the Ising model is exactly integrable. By using the tools of conformal perturbation theory~\cite{Guida:1995kc}, the spin-spin correlator can be obtained in principle for any spin model characterized by a second order phase transition (not only in two dimensions but also for three-dimensional universality classes,  see for instance refs.~\cite{Caselle:2015csa, Caselle:2016mww}) and thus the Svetitsky-Yaffe mapping can be used also for spin models that are not exactly integrable, like for instance the mapping between the (3+1) dimensional $\SU(2)$ LGT and the three dimensional Ising model discussed in \cite{Caselle:2019tiv}.


\providecommand{\href}[2]{#2}

\end{document}